\documentclass[twocolumn,superscriptaddress,amsmath,amssymb,floatfix,aps,prl]{revtex4}

\usepackage{graphicx}
\usepackage{dcolumn}
\usepackage{bm}
\usepackage{color}
\usepackage{float}
\usepackage{xr}
\usepackage{sidecap}
\usepackage{epstopdf}
\usepackage[UKenglish]{babel}
\begin{document}

\title{Topological Phononic Crystals with One-Way Elastic Edge Waves}

\author{Pai Wang}
\affiliation{School of Engineering and Applied Sciences, Harvard University, Cambridge, Massachusetts 02138, USA}
\author{Ling Lu}
\affiliation{Department of Physics, MIT, Cambridge, Massachusetts 02139, USA}
\author{Katia Bertoldi}
\thanks{Corresponding author. bertoldi@seas.harvard.edu}
\affiliation{School of Engineering and Applied Sciences, Harvard University, Cambridge, Massachusetts 02138, USA}
\affiliation{Kavli Institute, Harvard University, Cambridge, Massachusetts 02138, USA}

\date{\today}

\begin{abstract}
We report a new type of phononic crystals with topologically non-trivial bandgaps for both longitudinal and transverse polarizations, resulting in protected one-way elastic edge waves.
In our design, gyroscopic inertial effects are used to break the time-reversal symmetry and realize the phononic analogue of the electronic quantum Hall effect.
We investigate the response of both hexagonal and square gyroscopic lattices and observe bulk Chern number of 1 and 2, indicating that these structures support single and multi-mode edge elastic waves immune to back-scattering. These robust one-way phononic waveguides could potentially lead to the design of a novel class of surface wave devices that are widely used in electronics, telecommunication and acoustic imaging.
\end{abstract}

\pacs{Valid PACS appear here}


\maketitle


Topological states in electronic materials, including the quantum Hall effect~\cite{Klitzing1986} and topological insulators~\cite{Hasan2010,Qi2011}, have inspired a number of recent developments in photonics~\cite{Haldane2008,Lu2014}, phononics~\cite{Prodan2009,Zhang2010,Susstrunk2015,Xiao2015Geo,Xiao2015Syn} and mechanical metamaterials ~\cite{Sun2012,Kane2014,Paulose2015,Chen2014}.
In particular, in analogy to the quantum anomalous Hall effect~\cite{Haldane1988}, one-way waveguides in two-dimensional photonic systems have been realized by breaking time-reversal symmetry~\cite{Wang2008,Wang2009,Skirlo2014}. Moreover, very recently unidirectional edge channels have also been proposed for scalar acoustic waves   by rotating fluids~\cite{Fleury2014,Yang2015} and for elastic waves by Coriolis force in a non-inertial reference frame~\cite{Wang2014Coriolis}. However, the latter system is very difficult to implement in solid state devices.

Here, we introduce gyroscopic phononic crystals, where each lattice site is coupled with a spinning gyroscope that breaks time-reversal symmetry in a well-controlled manner.
In both hexagonal and square lattices, gyroscopic coupling opens bandgaps characterized by  Chern numbers of 1 and 2.
As a result, at the edge of these lattices both single-mode and multi-mode one-way elastic waves are observed to propagate around arbitrary defects without backscattering.

\begin{figure}
	\begin{center}
\includegraphics[width=1.0\columnwidth]{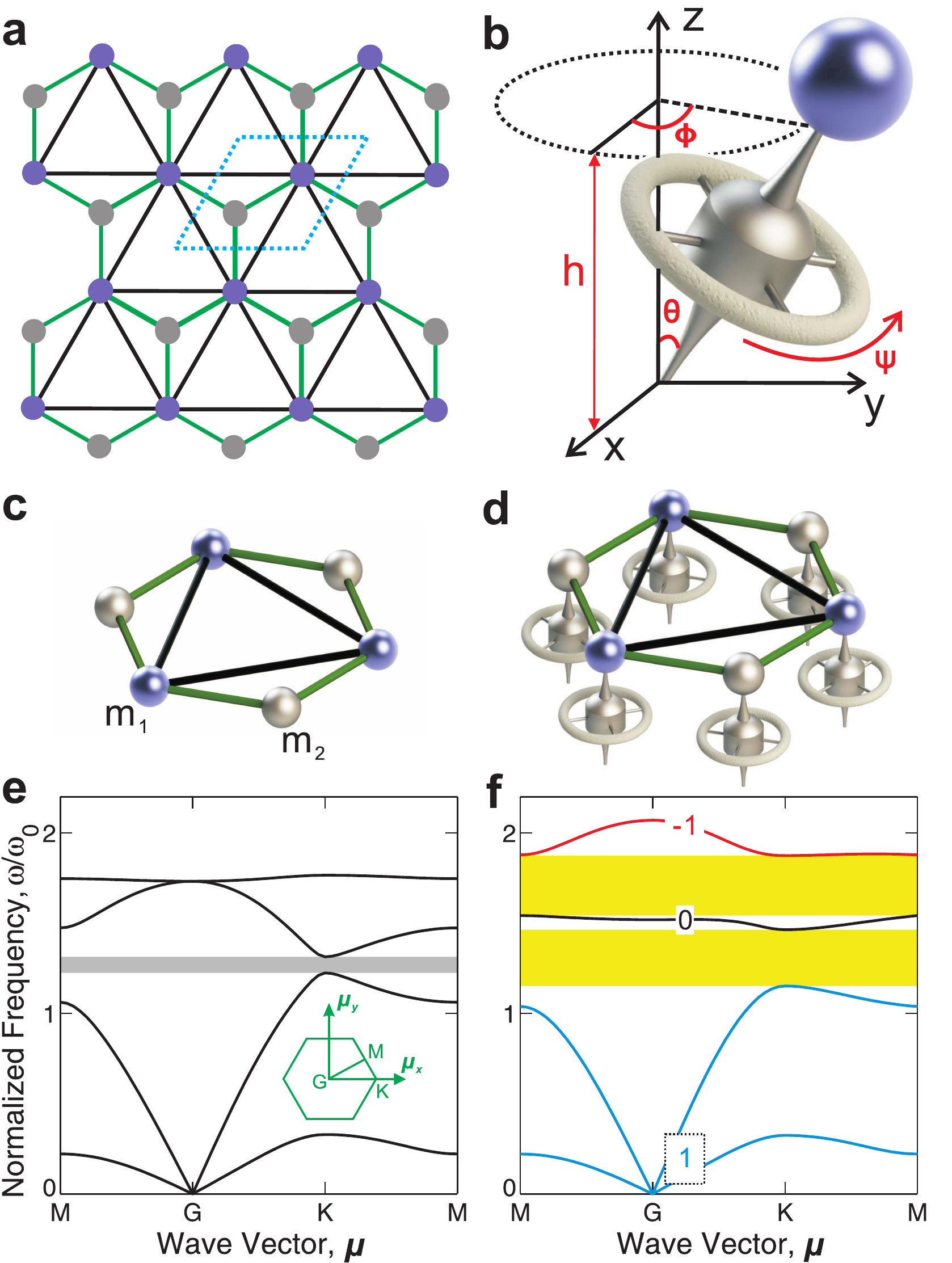}
\caption{\label{FIG:Design}\textbf{Ordinary and Gyroscopic Phononic Crystals:}
\textbf{(a)} Schematic of the hexagonal lattice. The purple and grey sphere represent concentrated masses  $m_1$ and $m_2$, respectively. The green and black straight rods represent mass-less linear springs with stiffness $k_1$ and $k_2 = k_1/20$, respectively. The dashed cell is the primitive cell of the lattice.
\textbf{(b)} Schematic of a gyroscope with the top tip pinned to a mass in the lattice. \textbf{(c)} Unit cell for the ordinary (non-gyroscopic) phononic crystal.
\textbf{(d)} Unit cell for the gyroscopic phononic crystal. \textbf{(e)} Band structure of the ordinary (non-gyroscopic) phononic crystal ($\alpha_1=\alpha_2=0$). The inset is the Brillouin zone.
\textbf{(f)} Band structure of the gyroscopic phononic crystal ($\alpha_1=\alpha_2=0.3m_1$) with the Chern numbers labeled on the bulk bands. The frequencies are normalized by $\omega_0 = \sqrt{k_1/m_1}$.}
	\end{center}
	\vspace{-15pt}
\end{figure}

To start, we consider a hexagonal phononic crystal with equal masses ($m_2 = m_1$) connected by linear springs (green and black rods in Figs. \ref{FIG:Design}a, \ref{FIG:Design}b and \ref{FIG:Design}c).
The resulting unit cell has four degrees of freedom specified by the displacements of $m_1$ and $m_2$ ($\mathbf{U}=[u_x^{m_1},\,u_y^{m_1},\,u_x^{m_2},\,u_y^{m_2}]$).
Consequently, there are a total of four bands in the band structure~(Fig. \ref{FIG:Design}e). Note that this is the minimal number of bands required to open a complete bandgap, since the first two elastic dispersions are pinned at zero frequency.
The phononic band structures are calculated by solving the dispersion equation~\cite{HusseinReview2014}
\begin{equation}
\label{EQN:DISPERSION}
\left[\mathbf{K}(\boldsymbol{\mu})-\omega^2 \mathbf{M} \right]\mathbf{U} =  \mathbf{0}
\end{equation}
for wave vectors $\boldsymbol{\mu}$ within the first Brillioun zone. Here, $\omega$ denotes the angular frequency of the propagating wave and $\mathbf{M} = \text{diag}\{m_1,\, m_1,\, m_2,\, m_2\}$  is the mass matrix. Moreover,
$\mathbf{K}$ is the 4$\times$4 stiffness matrix as a function of the Bloch wave vector $\boldsymbol{\mu}$.
The band structure of this simple lattice is shown in Fig. \ref{FIG:Design}e and has a quadratic degeneracy between the third and fourth bands at the center of the Brillouin zone and a complete gap between the second and third bands due to the lack of inversion symmetry. However, this gap is topologically trivial, since time-reversal symmetry is not broken and the Chern numbers of the bands are all zero.

In order to break time-reversal symmetry, we introduce  gyroscopic coupling~\cite{Milton2007,Brun2012,Carta2014} and attach each mass in the lattice to the tip of the rotational axis of a gyroscope, as shown in Fig. \ref{FIG:Design}d. Note that the other tip of the gyroscope is pinned to the ground to prevent any translational motion, while allowing for free rotations.  Because of the small-amplitude in-plane waves propagating in the phononic lattice,  the magnitude of the tip displacement of the each gyroscope is given by
\begin{equation}
U_{tip} =  h \sin{\theta} \approx h \theta = h \Theta  e^{i\omega t} \quad \mbox{for} \quad |\Theta| \ll 1,
\end{equation}
where $h$ and $\theta$ denote the height and nutation angle of the gyroscope (Fig. \ref{FIG:Design}b) and $\Theta$ is the amplitude of the harmonic change in $\theta$.
Interestingly, the coupling between the masses in the lattices and the gyroscopes induces an in-plane gyroscopic inertial force perpendicular to the direction of $U_{tip}$ ~\cite{Brun2012,Goldstein2002,Support}:
\begin{equation}
F_g = \pm i \omega^2 \alpha U_{tip},
\end{equation}
where $\alpha$ is the spinner constant that characterizes the strength of the rotational coupling between two independent inertias in the 2D plane.  As a result, the mass matrix in Eq. (\ref{EQN:DISPERSION}) becomes,
\begin{equation}
\label{EQN:MASS_MODIFIED}
\mathbf{\tilde{M}} = \mathbf{M} +
 \begin{pmatrix}
  0          & i\alpha_1 & 0         & 0 \\
  -i\alpha_1 &0          & 0         & 0 \\
  0          & 0         & 0         & i\alpha_2  \\
  0          & 0         &-i\alpha_2 & 0
 \end{pmatrix},
\end{equation}
where $\alpha_1$ and $\alpha_2$ denote the spinner constants of the gyroscopes attached to $m_1$ and $m_2$, respectively. We note that the imaginary nature of the gyroscopic inertial effect indicates directional phase shifts with respect to the tip displacements, which breaks time-reversal symmetry.

We now consider a gyroscopic hexagonal lattice with $\alpha_2 = \alpha_1 = 0.3 m_1$ and show its band structure  in Fig. \ref{FIG:Design}f. Comparing this to the original band structure reported in Fig. \ref{FIG:Design}e,
we observe that the original quadratic degeneracy, between the third and fourth bands, is opened into a full band gap.
Moreover, we also find that the original bandgap between the second and third bands first closes and then reopens as we gradually increase the magnitude of $\alpha_1$ and $\alpha_2$. In particular,  for $\alpha_1=\alpha_2=0.07m_1$  the gap is closed and a pair of Dirac cones at $K$ points emerges (see Fig. S1 for details). Importantly,  for $\alpha_2=\alpha_1=0.3m_1$ (Fig. \ref{FIG:Design}f) both bandgaps are topologically-nontrivial as highlighted by the non-zero Chern numbers labeled on the bands (the calculations conducted to compute these topological invariants are detailed in the Supplementary Materials~\cite{Support}). Therefore, in the frequency ranges of these nontrivial bandgaps, we expect gapless one-way edge states, whose number is dictated by the sum of Chern numbers below the bandgap.

\begin{figure}[t]
	\begin{center}
\includegraphics[width=1.0\columnwidth]{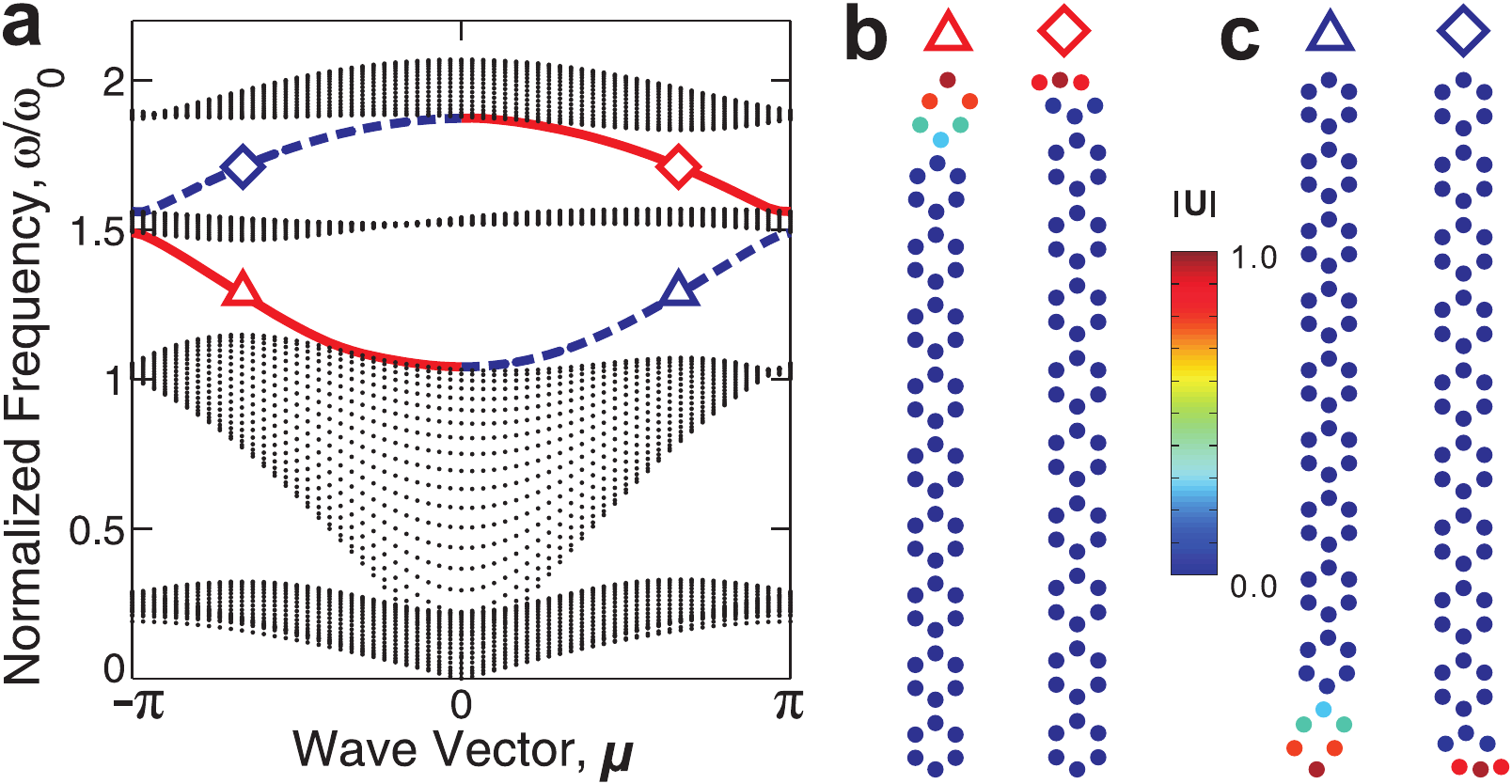}
\caption{\label{FIG:Edge}\textbf{Edge modes in Gyroscopic Phononic Crystal:} \textbf{(a)} 1D band structure showing bulk bands (black dots) and edge bands (colored lines). Red solid lines represent edge modes bound to the top boundary, while blue dashed lines represent edge modes bound to the bottom boundary. \textbf{(b)} Modal displacement fields of top edge states with negative group velocities. \textbf{(c)} Modal displacement fields of bottom edge states with positive group velocities}
	\end{center}
	\vspace{-15pt}
\end{figure}

To verify the existence of such one-way edge states, we perform one-dimensional (1D) Bloch wave analyses on a supercell comprising $20 \times 1$ unit cells, assuming free boundary conditions for the top and bottom edges. In full agreement with the bulk Chern numbers, the band structure of the supercell shows one one-way edge mode on each edge in both bandgaps.
For modes bound to the top edge~(Fig. \ref{FIG:Edge}b), the propagation can only assume negative group velocities (red solid lines with negative slope in Fig. \ref{FIG:Edge}a).
On the other hand, the modes bound to the bottom edge~(Fig. \ref{FIG:Edge}c) possess positive group velocities (blue dashed lines with positive slope in Fig. \ref{FIG:Edge}a).
Since these edge modes are in the gap frequency range where no bulk modes may exist, they cannot scatter into the bulk of the phononic crystal. Furthermore, their  uni-directional group velocities guarantee the absence of any back scattering and result in the topologically protected one-way propagation of vibration energy.\\

\begin{figure}[h]
	\begin{center}
\includegraphics[width=1.0\columnwidth]{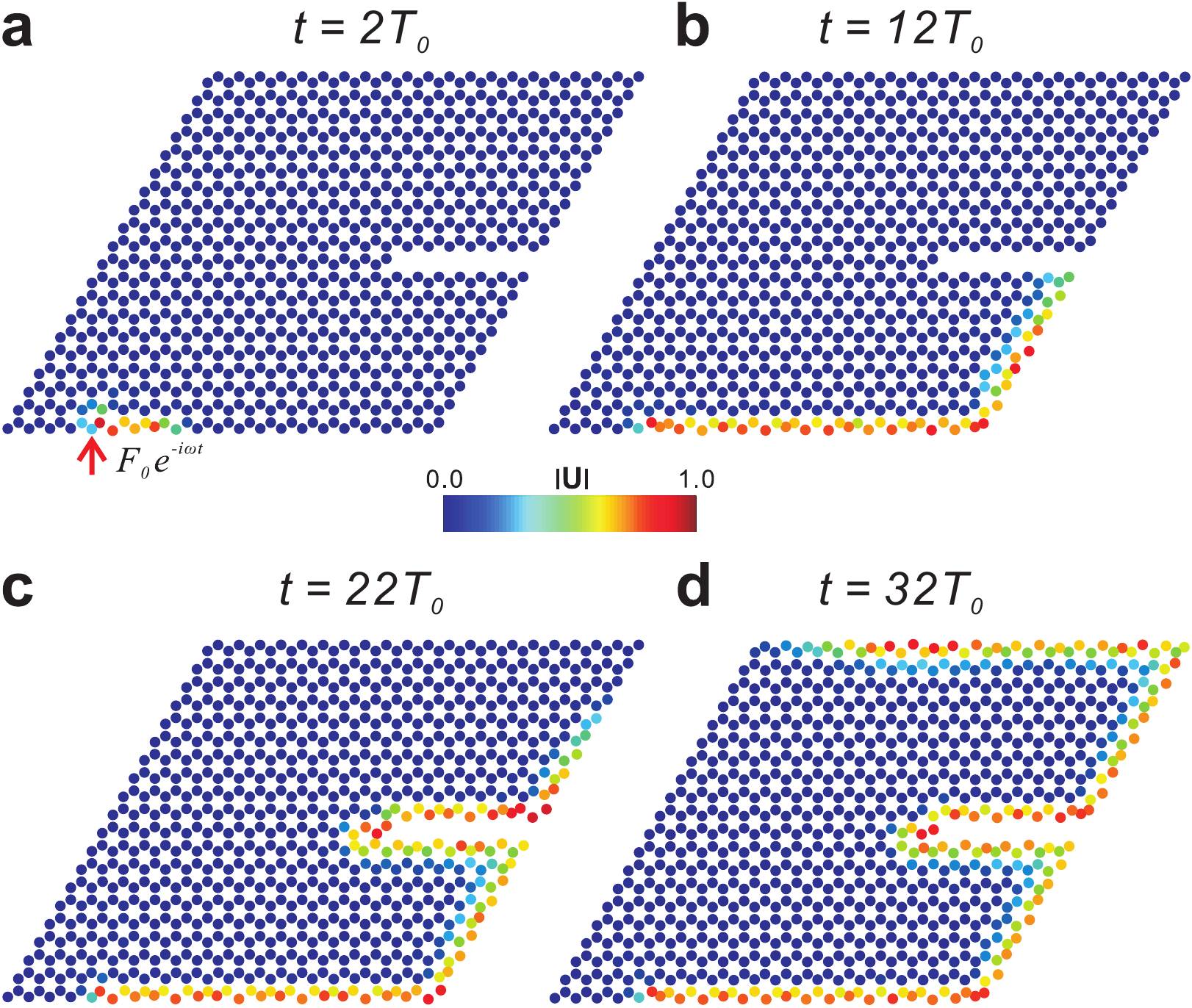}
\caption{\label{FIG:Transient}\textbf{Transient Response} of a gyroscopic phononic crystal consisting of 20 $\times$ 20 unit cells with a line defect on the right boundary: Snapshots of the displacement field at \textbf{(a)} $t=2T_0$, \textbf{(b)} $t=12T_0$, \textbf{(c)} $t=22T_0$ and \textbf{(d)} $t=32T_0$, where $T_0 = \sqrt{m_1/k_1}$ is the characteristic time scale of the system. Starting from $t=0$, a time-harmonic excitation force $\mathbf{F}(t)=[F_x(t),\,F_y(t)]=[1,\,1] F_0e^{-i \omega t}$ is prescribed at the site indicated by the red arrow\color{black}.}
	\end{center}
	\vspace{-15pt}
\end{figure}

To show the robustness of these edge states, we conduct transient analysis on a finite sample comprising $20 \times 20$ unit cells with a line defect on the right boundary created by removing twelve masses and the springs  connected to them~(Fig. \ref{FIG:Transient}a). A harmonic force excitation, $F_0 \text{e}^{-i\omega t}$, is prescribed at a mass site on the bottom boundary (red arrow in Fig. \ref{FIG:Transient}a) with frequency  within the bulk bandgap between the second and third bands ($\omega/\omega_0= 1.3$). In Fig. \ref{FIG:Transient} we plot snapshots of the velocity field at different time instances, $t/T_0=$ 2, 12, 22 and 32, where $T_0 = \sqrt{m_1/k_1}$ is the characteristic time scale of the system. Remarkably, because of their topological protection, the edge modes circumvent both the sharp corner and the line defect without any reflection.

\begin{figure}[h]
	\begin{center}
\includegraphics[width=1.0\columnwidth]{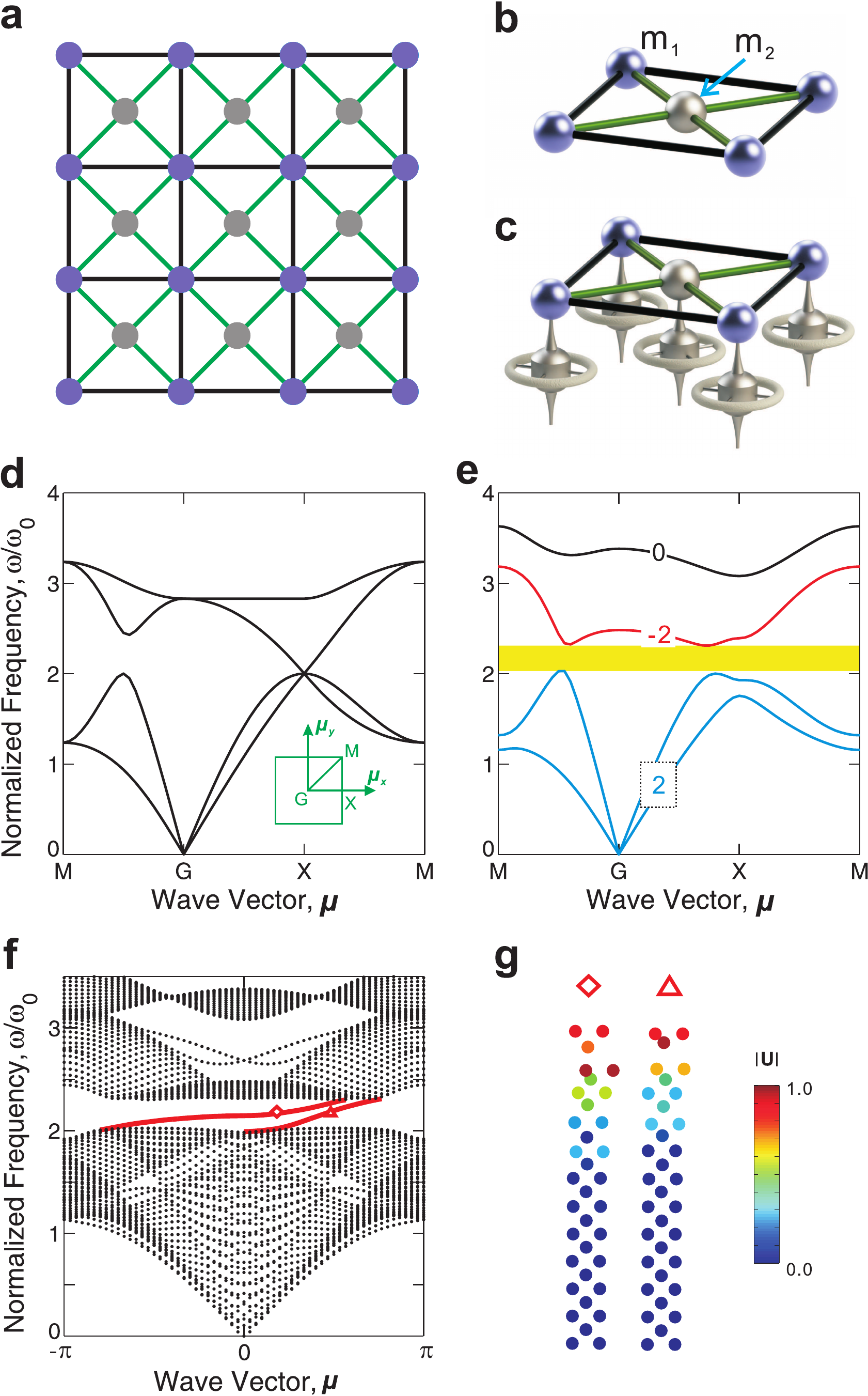}
\caption{\label{FIG:SQ}\textbf{Square Lattice results:} \textbf{(a)} Schematic of the square lattice. The purple and grey sphere represent concentrated masses  $m_1$ and $m_2$, respectively. The black and green straight rods represent mass-less linear springs with stiffness $k_1$ and $k_2 = 2k_1$, respectively. \textbf{(b)} Unit cell for the ordinary (non-gyroscopic) phononic crystal. \textbf{(c)} Unit cell for the gyroscopic phononic crystal.
\textbf{(d)} Band structure of the ordinary (non-gyroscopic) phononic crystal ($\alpha_1 = \alpha_2 =0$). The insets is the Brillouin zone.
\textbf{(e)} Band structure of the gyroscopic phononic crystal ($\alpha_1$ = $\alpha_2=0.3 m_1$) with the Chern numbers labeled on the bulk bands. Frequencies are normalized by $\omega_0 = \sqrt{k_1/m_1}$.
\textbf{(f)} 1D band structure showing bulk bands (black dots) and topological edge bands (red solid lines \color{black}for edge modes bound to the top boundary. Note that the edge bands that are bound to the bottom boundary are not shown here. \textbf{(g)} Modal displacement fields of the edge states shown in \textbf{(f)}.}
	\end{center}
	\vspace{-15pt}
\end{figure}

Next, we investigate the effect of the lattice geometry and start with an ordinary square phononic crystal with masses $m_1$ connected by springs with elastic constant $k_1$. To make the lattice statically stable,  we add an additional mass $m_2=m_1$ at the center of each unit cell and connect it to its four adjacent $m_1$ masses by springs with elastic constant $k_2=2k_1$ (See Figs. \ref{FIG:SQ}a and \ref{FIG:SQ}b).
The band structure for this lattice (shown in Fig. \ref{FIG:SQ}d) contains a pair of three-fold linear degeneracy among the first, second and third bands at the $X$ points of the Brillouin zone.
Note that this type of degeneracy, consisting of a locally flat band and a Dirac point, is known as the ``accidental Dirac point'' ~\cite{liu2011dirac}. Interestingly, while previously  this degeneracy has been found to be very sensitive to the system  parameters ~\cite{liu2011dirac}, in our lattice
it robustly appears  at the $X$ points when $m_1=m_2$.
Upon the introduction of gyroscopic inertial effects ($\alpha_2 = \alpha_1 = 0.3\,m_1$), these three-fold degenerate points are lifted and a gap is created between the second and third bands.
The Chern numbers of the two bulk bands below the gap is two, predicting the existence of two topological edge states.
In Figs. \ref{FIG:SQ}f and \ref{FIG:SQ}g, we plot the band structure of the corresponding  20$\times$1 supercell, highlighting the two one-way edge modes and their modal displacement fields.


To summarize, we demonstrated that gyroscopic phononic crystals can support topologically non-trivial gaps, within which the edge states are unidirectional and immune to back-scattering.
Moreover, transient analysis confirmed that the propagation of such topological edge waves is robust against large defect and sharp corners.
We note that phononic crystals~\cite{Sigalas1993,Kushwaha1993} and acoustic metamaterials~\cite{Liu2000,Fang2006,Ding2007,Lai2011,Wang2014Tunable} that enable manipulation and control of elastic waves have received significant interest in recent years~\cite{MaldovanBook2009,HusseinReview2014}, not only because of their rich physics, but also for their broad range of applications~\cite{Ruzzene2003,Fleck2006,Vasseur2007,Kafesaki2000,Pennec2004,Mei2012,Brule2014,Chen2004,Christensen2007,Li2009,Bigoni2013,Maldovan2013,Hussein2014}. Interestingly, the edge wave modes in phononic crystals are important in many scenarios~\cite{Torres1999,Torrent2012,Li2014}, including vibration control~\cite{Torrent2013} and acoustic imaging~\cite{Christensen2007}. However, most of reported studies have focused on topologically trivial surface waves that can be easily scattered or localized by defects~\cite{Torres1999}. Therefore, the work reported here could open new avenues for the design of phononic devices  with special properties and functionalities on edges, surfaces and interfaces.

\begin{acknowledgments}
This work has been supported by NSF through grants  CMMI-1120724, CMMI-1149456 (CAREER) and the Materials Research Science and Engineering Center (DMR-1420570). K.B. acknowledges start-up funds from the Harvard School of Engineering and Applied Sciences and the support of the Kavli Institute and Wyss Institute at Harvard University.
L.L. was supported in part by U.S.A.R.O. through the ISN under Contract No. W911NF-13-D-0001, in part by the MRSEC Program of the NSF under Award No. DMR-1419807 and in part by the MIT S3TEC EFRC of DOE under Grant No. DE-SC0001299.
The authors are also grateful to Scott Skirlo, Timothy H. Hsieh and Dr. Filippo Casadei for ispirational discussions and to Dr. Farhad Javid and Xin You for their support with graphics.
\end{acknowledgments}

\bibliography{Gyroscope}


%
%

\end{document}